\documentclass[english,reprint,showpacs,preprintnumbers,nofootinbib,amsmath,amssymb,prl]{revtex4-1}
\usepackage[latin9]{inputenc}
\usepackage{verbatim}
\usepackage{amsmath}
\usepackage{amssymb}
\usepackage{graphicx}

\makeatletter
\@ifundefined{textcolor}{}
{%
\definecolor{BLACK}{gray}{0}
\definecolor{WHITE}{gray}{1}
\definecolor{RED}{rgb}{1,0,0}
\definecolor{GREEN}{rgb}{0,1,0}
\definecolor{BLUE}{rgb}{0,0,1}
\definecolor{CYAN}{cmyk}{1,0,0,0}
\definecolor{MAGENTA}{cmyk}{0,1,0,0}
\definecolor{YELLOW}{cmyk}{0,0,1,0}
}

\@ifundefined{showcaptionsetup}{}{%
 \PassOptionsToPackage{caption=false}{subfig}}
\usepackage{subfig}
\makeatother

\usepackage{babel}
\begin{document}

\title{A maximum entropy approach to separating noise from signal in bimodal
affiliation networks}

\author{Navid Dianati}

\thanks{n.dianatimaleki@neu.edu}

\affiliation{The Lazer Lab, Northeastern University, Boston Massachusetts.}

\affiliation{Institute for Quantitative Social Sciences, Harvard University, Cambridge
Massachusetts.}
\begin{abstract}
In practice, many empirical networks, including co-authorship and
collocation networks are unimodal projections of a bipartite data
structure where one layer represents entities, the second layer consists
of a number of sets representing affiliations, attributes, groups,
etc., and an interlayer link indicates membership of an entity in
a set. The edge weight in the unimodal projection, which we refer
to as a \textit{co-occurrence network}, counts the number of sets
to which both end-nodes are linked. Interpreting such dense networks
requires statistical analysis that takes into account the bipartite
structure of the underlying data. Here	 we develop a statistical significance
metric for such networks based on a maximum entropy null model which
preserves both the frequency sequence of the individuals/entities
and the size sequence of the sets. Solving the maximum entropy problem
is reduced to solving a system of nonlinear equations for which fast
algorithms exist, thus eliminating the need for expensive Monte-Carlo
sampling techniques. We use this metric to prune and visualize a number
of empirical networks.
\end{abstract}
\maketitle

\section{Introduction}

Many integer weighted graphs derived from empirical data are so-called
\textit{co-occurrence }graphs: an edge weight counts the number of
times the two end nodes where observed to share a property. Most abstractly,
this shared property can be modeled as membership in some unordered
set. For instance, membership in the same team or group, affiliation
with an institution, shared physical attributes, or words appearing
in the same document. Such networks have been studied in various contexts
including co-attendance in social events \cite{davis2009DeepSouth},
networks of co-starring actors \cite{watts1998collective} congressional
bill co-sponsorship networks \cite{fowler2006connecting,fowler2006legislative}. 

Formally, given a set $S=\{s_{1},s_{2},\cdots,s_{m}\}$ of symbols
or entities, the data consists of an arbitrary number of subsets of
$S:$
\begin{equation}
D=\left\{ u_{j}\right\} _{j=1}^{n},\,\,\,\,\,u_{j}\subset S\,\,\,\,j=1,\cdots n.
\end{equation}
In its simplest form, a given entry $u_{j}$ is simply an unordered
set defining a symmetric relationship between every pair of its elements.
A weighted graph may then be defined with vertex set $V\equiv S$
where a weighted edge between two nodes counts the number of subsets
$u_{j}$ containing both nodes. In the context of natural language
processing, the subsets $u_{j}$ are commonly referred to as \textit{documents
}and their elements as \textit{words }or \textit{symbols. }%
\begin{comment}
\textbf{(verify)}\textit{ }
\end{comment}
The set $S$ is sometimes referred to as the \textit{Lexicon. }We
will use this terminology in the rest of this paper. 

Depending on the nature of the data, more specialized ways of constructing
a graph may be desirable. For instance, a document may contain an
internal order in which case different pairs of symbols within the
document may need to be assigned different weights. In this paper
we will consider the most generic case of unordered sets and homogeneous
co-occurrence weights.

\begin{figure}
\noindent \begin{centering}
\includegraphics[width=0.8\columnwidth]{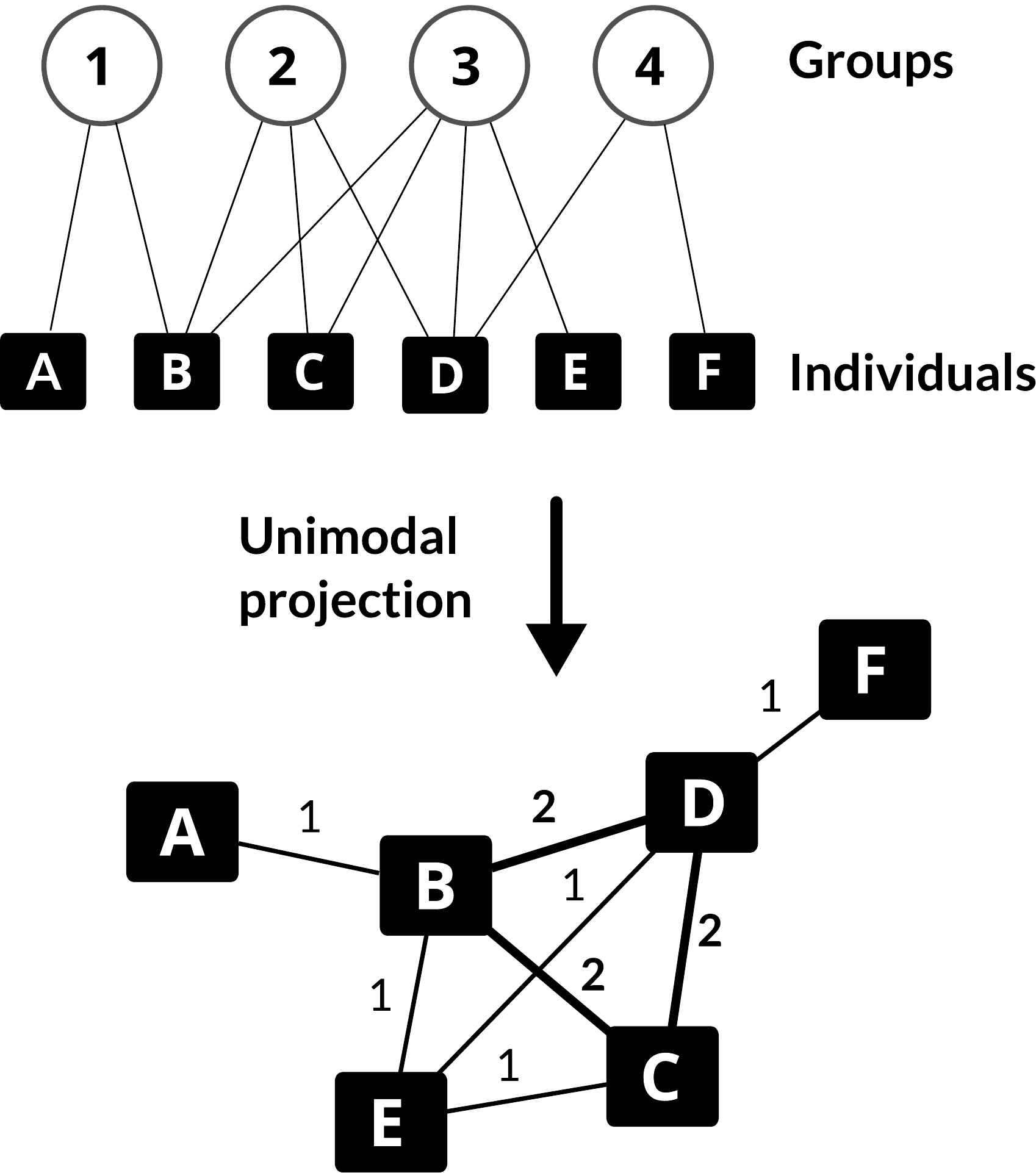}
\par\end{centering}

\caption{}

\noindent \centering{}\caption{{\small{}Example of a co-occurrence network compiled from a bimodal
entity-affiliation graph.\label{fig:Example-of-projection}}}
\end{figure}

The data can be abstracted as a bipartite network where the vertex
set for one layer consists of the set $S$ of all symbols and the
vertex set of the second layer is the set of all sets $u_{\alpha},$
$\alpha=1,\cdots n.$ An edge between a symbol $s_{i}$ and a set
$u_{\alpha}$ denotes the relation $s_{i}\in u_{\alpha}.$ Let $g_{\alpha}=\left|u_{\alpha}\right|,\,\alpha=1,2,\cdots n$
denote the size of the set $u_{\alpha},$ and $f_{i},\,i=1,2,\cdots m$
the frequency of the symbol $s_{i}$ in the entire dataset. In the
bipartite graph, these two sequences are then simply the degrees of
the corresponding nodes in the first and second layers respectively
and we trivially have $\sum_{i}f_{i}=\sum_{\alpha}g_{\alpha}=N$.
The co-occurrence network is then a weighted \textit{projection }of
this bipartite graph onto the layer consisting of the symbols or entities.
See figure \ref{fig:Example-of-projection} for an example.

The question of most practical interest is how one can extract statistically
meaningful substructures in the co-occurrence network. These structures---which
are believed to be obscured by an abundance of noisy edges---are sometimes
called the \textit{backbone }of the network and the removal of insignificant
edges in the hope of uncovering them is referred to as \textit{pruning.
}Most commonly, pruning is performed by \textit{weight thresholding,
}i.e., removing the edges with weights below a desired threshold from
the graph. This is a naive approach as it results in the loss of the
multiscale structure of the graph. For natively unimodal networks,
other statistically inspired methods have been proposed including
the \textit{disparity filter} \cite{serrano_extracting_2009}, the
GLOSS filter \cite{radicchi_information_2011},\textit{ }and the \textit{marginal
likelihood filter }(MLF) \cite{dianati_unwinding_2016}. These methods
are statistically informed since they formulate generative null models
and then identify features in the observed network least expected
to have occurred due to pure chance according to the null model. Similar
methodologies have also been proposed for bimodal networks of the
kind we are concerned with in this paper, including the \textit{fixed
degree sequence model }(FDSM)\cite{Latapy200831} and \textit{stochastic
degree sequence model }(SDSM) \cite{neal_backbone_2014}. These latter
methods employ random null models that preserve the degree sequences
of the nodes in both layers (corresponding to the frequency sequence
of the symbols and the size sequence of the sets) with FDSM doing
so exactly and SDSM on average. 

In this paper we propose a random graph ensemble also based on the
same intuition as the SDSM---namely preserving the expectation value
of the full degree sequence of the graph---and its resulting significance
test. Our methodology differs from the SDSM in important ways. Firstly,
the SDSM generates realizations of the random graph ensemble by sampling
each possible edge in the bipartite graph according to a Bernoulli
process whose probability is determined by solving a regression model
such that the expectation value of each node's degree matches the
corresponding degree in the observed graph with reasonable precision.
While this randomization process generates an ensemble approximately
consistent with the desired constraint, it is not guaranteed to yield
the ``most random'' such ensemble. By contrast, in this paper we
compute an ensemble that is in fact guaranteed to be the most random
(i.e., the most unbiased) one satisfying the constraint, by solving
a maximum entropy problem. Secondly, the test statistics of the SDSM
are computed by sampling the graph ensemble and deriving empirical
null distributions for the co-occurrence edges based on the obtained
sample. The accuracy of the test statistics is thus critically dependent
on the sample size, making it computationally expensive to produce
reliable results. We, on the other hand, derive test statistics that
can be computed exactly, or otherwise with high precision without
the need to sample the ensemble.

\begin{comment}
\begin{figure}
\centering{}\includegraphics[width=0.6\columnwidth]{senate_110_cooccurrence_full.png}\caption{{\small{}An example of a dense co-occurrence network: the full US
senate bill cosponsorship network for the 110th congress. The network
consists of 102 nodes and 5151 weighted edges where the weight counts
the number of bills the two senators have cosponsored. Node colors
indicate party membership.} Data from \cite{fowler2006connecting,fowler2006legislative}.\label{fig:The-full-senate-graph}}
\end{figure}
\end{comment}

\section{Unweighted co-occurrence networks}

Let us focus on the case where the link between a symbol and a set
is unweighted, that is, a symbol either appears in a set or it doesn't.
We must formulate a randomization process whereby some set of meaningful
and presumably robust features of the observed graph are preserved
on average but the graph is randomized otherwise. We choose to preserve
the degree sequences of both layers, one corresponding to the frequencies
of the symbols throughout the data set, and the other corresponding
to the size sequence of the sets to which the symbols can be related
by membership. At first glance, this problem appears to be simply
a bipartite analogue of the \textit{Marginal Likelihood Filter }\cite{dianati_unwinding_2016}
where for a given unimodal, integer-weighted event-counting network,
a set of independent assignment events are distributed randomly between
all possible node pairs such that the degree sequence is preserved
on average. But the present problem is different for two reasons:
1) the inter-layer edges cannot be modified independently of one another
since the co-membership relation is transitive, and 2) randomly distributing
assignment events would allow for multiedges. We may then simply demand
that a given pair $s_{i},u_{\alpha}$ be connected with probability
$f_{i}g_{\alpha}/N$ which does indeed lead to the correct expectation
value for both degrees. However, there is no guarantee that this quantity
is even a probability. Instead, we derive a \textit{maximum entropy
ensemble }with the desired constraints, hoping to be able to compute
the marginal probability distributions for all $(i,\alpha)$ edges,
leading to a simple marginal significance test similar to \cite{dianati_unwinding_2016}.

\begin{figure*}
\centering{}\subfloat[]{\includegraphics[height=3.9cm]{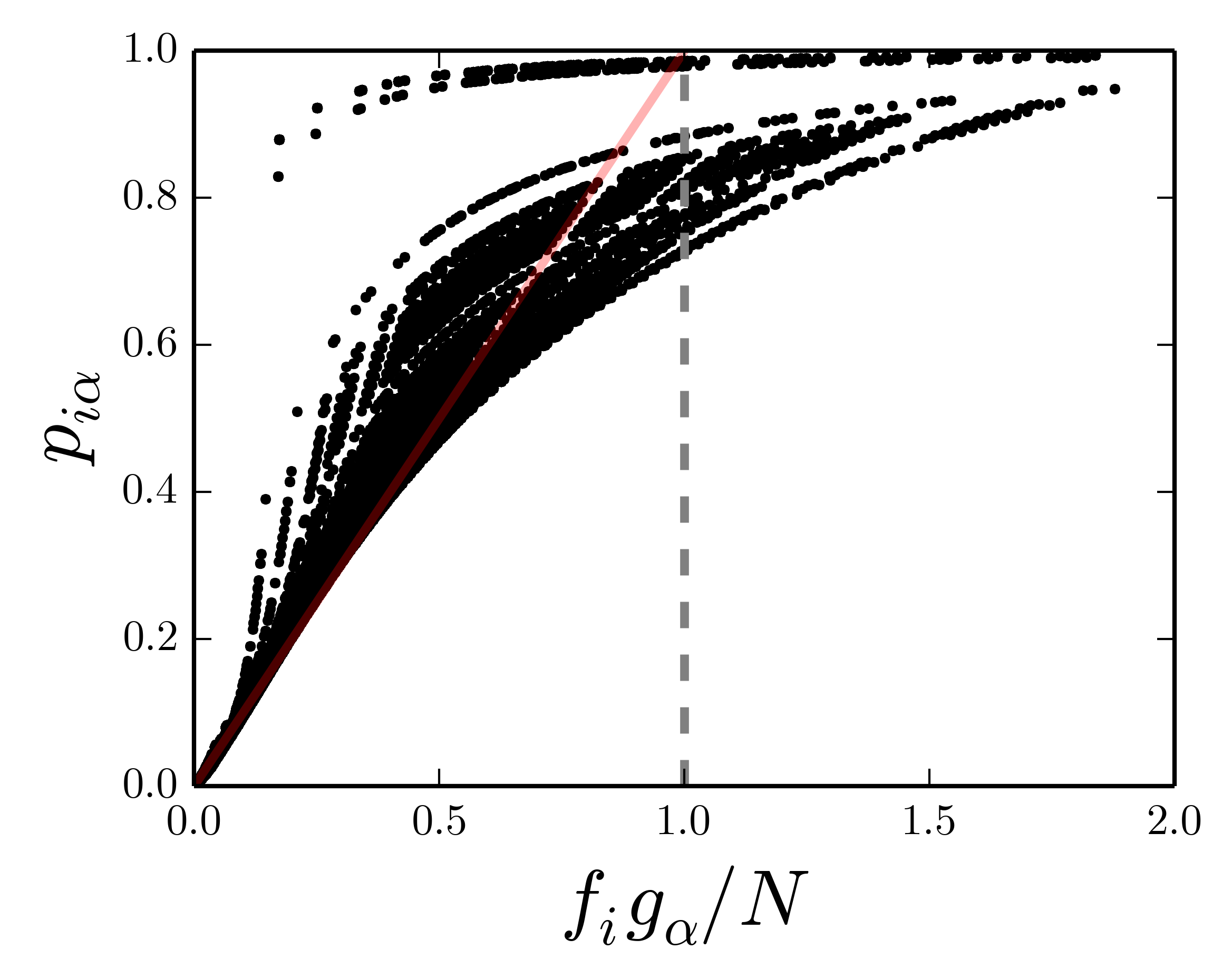}}\subfloat[]{\includegraphics[height=3.9cm]{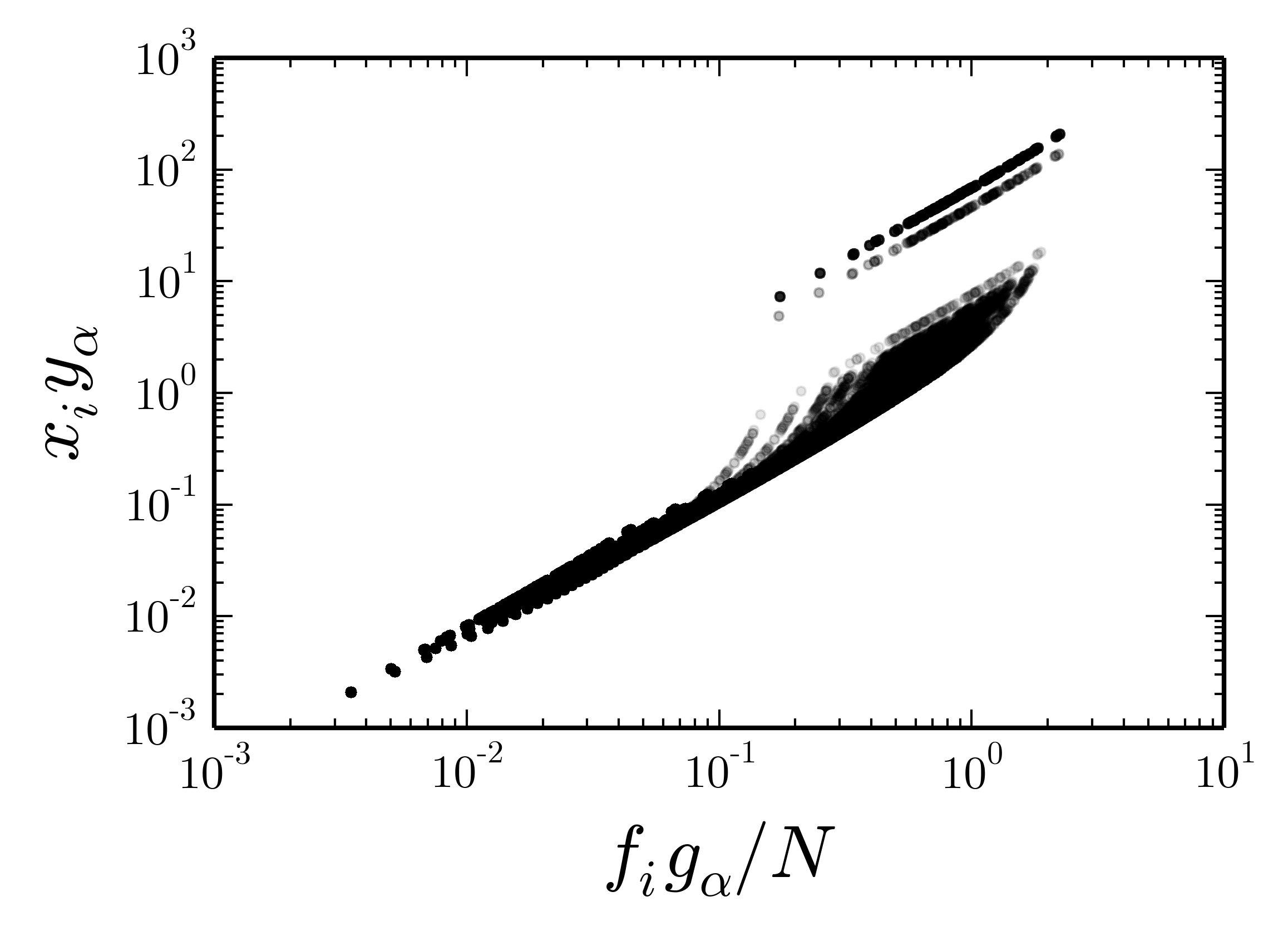}}\subfloat[]{\includegraphics[height=3.9cm]{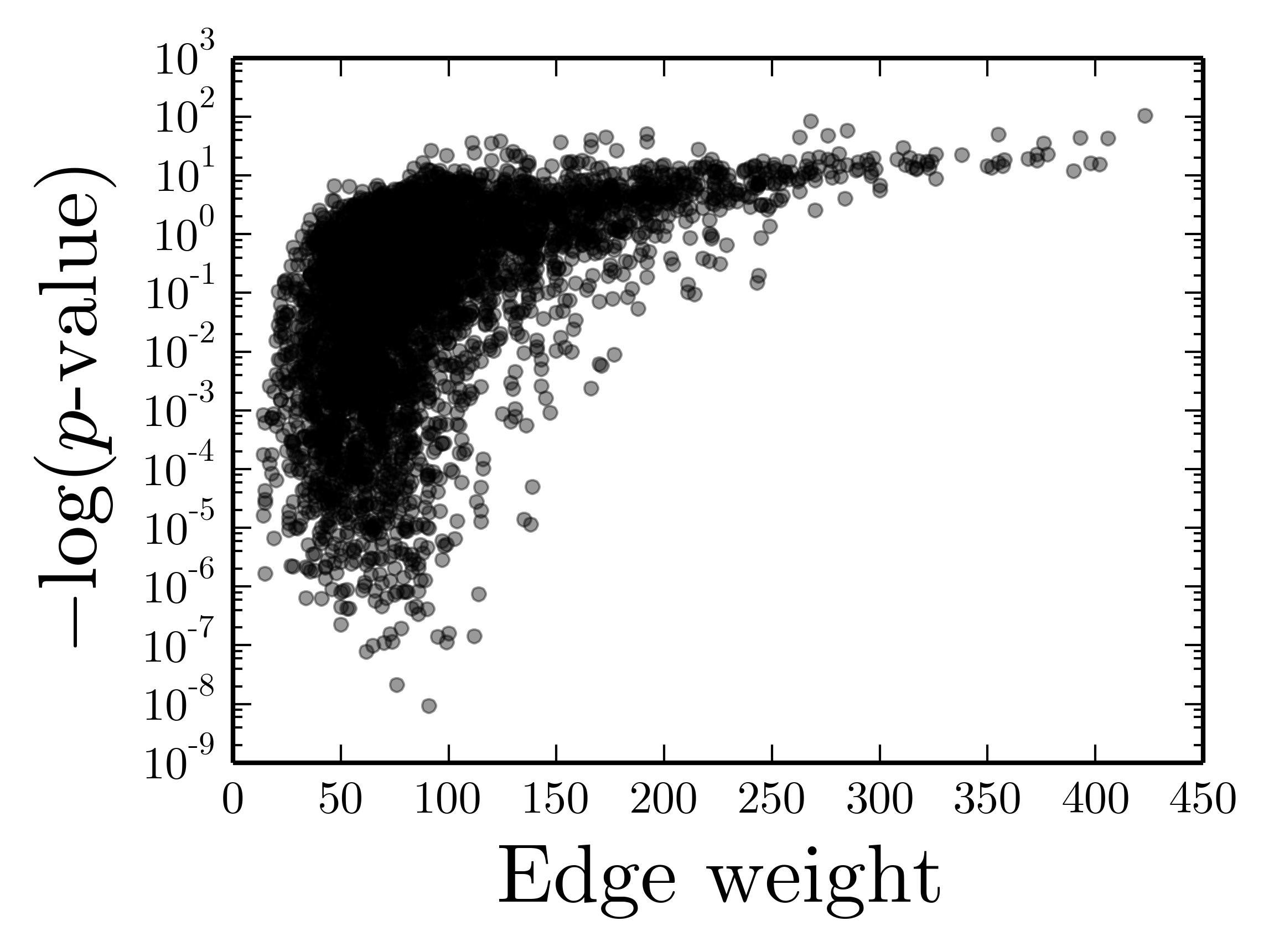}}\caption{(a) The numerically computed connection probabilities between the
two layers for the senate co-sponsorship network of the 110th US congress.
Plotted against the ``naive'' probability. Note the highly nonlinear
dependence. (b) Numerically solved $x_{i}y_{\alpha}$ as a function
of the first order guess. In both plots, a cluster of points stand
out, far above the bulk. These correspond to pairs with $g_{\alpha}\simeq n,$
i.e., sets that are connected to almost every entity, leading to near
certain expected connectivity according to the null model. (c) \textit{p-}value
vs weight for edges in the co-occurrence graph. \label{fig:solution-numerical}}
\end{figure*}

Let us derive a maximum entropy graph ensemble that preserves both
the set sizes $g_{\alpha}$ and symbol frequencies $f_{i}$ on average.
The probability distribution for this ensemble is given by an exponential
where our $m+n$ linear constraints $\left\langle \sum_{\alpha}\sigma_{i\alpha}\right\rangle =f_{i},\,\,i=1,2,\cdots m$
and $\left\langle \sum_{i}\sigma_{i\alpha}\right\rangle =g_{\alpha},\,\,\alpha=1,2,\cdots n$
are enforced by Lagrange multipliers $\lambda_{i},\,\,i=1,2,\cdots m$
and $\gamma_{\alpha},\,\,\alpha=1,2,\cdots n.$
\begin{eqnarray}
P(G) & \sim & \exp\left[\sum_{i}\lambda_{i}\sum_{\alpha}\sigma_{i\alpha}+\sum_{\alpha}\gamma_{\alpha}\sum_{i}\sigma_{i\alpha}\right]\nonumber \\
 & = & \exp\left[\sum_{i,\alpha}\left(\lambda_{i}+\gamma_{\alpha}\right)\sigma_{i\alpha}\right]
\end{eqnarray}
where $\sigma_{i\alpha}$ is either zero or one, indicating whether
nodes $i$ and $\alpha$ from the first and second layers respectively
are connected. Therefore, the partition function is given by
\begin{eqnarray}
Z & = & \sum_{\left\{ \sigma_{i\alpha}\right\} }e^{\sum_{i,\alpha}\left(\lambda_{i}+\gamma_{\alpha}\right)\sigma_{i\alpha}}\\
 & = & \sum_{\left\{ \sigma_{i\alpha}\right\} }\prod_{i,\alpha}e^{\left(\lambda_{i}+\gamma_{\alpha}\right)\sigma_{i\alpha}}\\
 & = & \prod_{i,\alpha}\left[1+e^{\left(\lambda_{i}+\gamma_{\alpha}\right)}\right].
\end{eqnarray}
Now we enforce the constraints and compute the Lagrange multipliers:
\begin{equation}
f_{j}=\frac{\partial\log Z}{\partial\lambda_{j}}\qquad\mbox{and}\qquad g_{\beta}=\frac{\partial\log Z}{\partial\gamma_{\beta}}.
\end{equation}
Thus,%
\begin{comment}
\begin{eqnarray}
f_{j} & = & \frac{\partial}{\partial\lambda_{j}}\sum_{i,\alpha}\log\left[1+e^{\left(\lambda_{i}+\gamma_{\alpha}\right)}\right]\\
 & = & \sum_{\alpha}\frac{e^{\lambda_{j}+\gamma_{\alpha}}}{1+e^{\lambda_{i}+\gamma_{\alpha}}}
\end{eqnarray}
\end{comment}
\begin{align}
f_{j} & =\sum_{\alpha}\frac{e^{\lambda_{j}+\gamma_{\alpha}}}{1+e^{\lambda_{i}+\gamma_{\alpha}}},\\
g_{\beta} & =\sum_{i}\frac{e^{\lambda_{i}+\gamma_{\beta}}}{1+e^{\lambda_{i}+\gamma_{\beta}}}.
\end{align}
Defining $x_{i}=e^{\lambda_{i}}$ and $y_{\alpha}=e^{\gamma_{\alpha}},$
our problem is reduced to the solution of the following system of
nonlinear equations:
\begin{eqnarray}
y_{\alpha} & = & g_{\alpha}/\sum_{i}\frac{x_{i}}{1+x_{i}y_{\alpha}},\alpha=1,\cdots,n\label{eq:logistics1}\\
x_{i} & = & f_{i}/\sum_{\alpha}\frac{y_{\alpha}}{1+x_{i}y_{\alpha}},i=1,\cdots,m.\label{eq:logistics2}
\end{eqnarray}
\begin{comment}
Viewed differently, we have a system of equations involving the \textit{logistic
function }$F(x)=\frac{e^{x}}{1+e^{x}}$:
\begin{align}
\sum_{\alpha}F(\lambda_{i}+\gamma_{\alpha}) & =f_{i}\\
\sum_{i}F(\lambda_{i}+\gamma_{\alpha}) & =g_{\alpha}
\end{align}
\end{comment}
Note that these equations are basically telling us that according
to the maximum entropy scheme, the ``occupation'' probability of
each of the possible edges between the first and second layers should
have a logistic form:
\begin{equation}
p_{i\alpha}=\frac{e^{\lambda_{i}+\gamma_{\alpha}}}{1+e^{\lambda_{i}+\gamma_{\alpha}}}=\frac{x_{i}y_{\alpha}}{1+x_{i}y_{a}}.
\end{equation}

\section{Solving the saddlepoint equations}

\begin{figure*}
\subfloat[]{\centering{}\includegraphics[width=0.7\columnwidth]{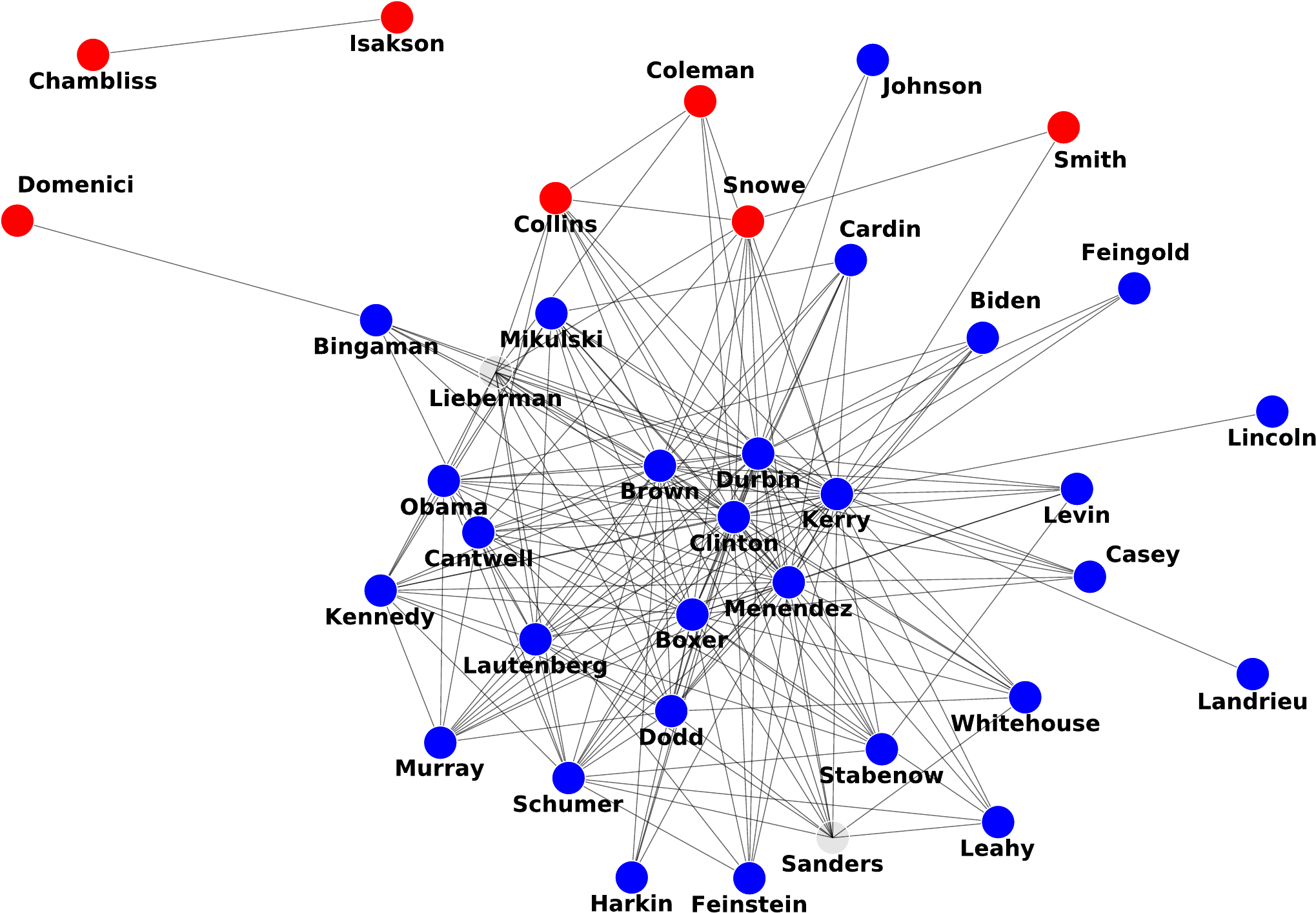}}\hfill{}\subfloat[]{\centering{}\includegraphics[width=1.32\columnwidth]{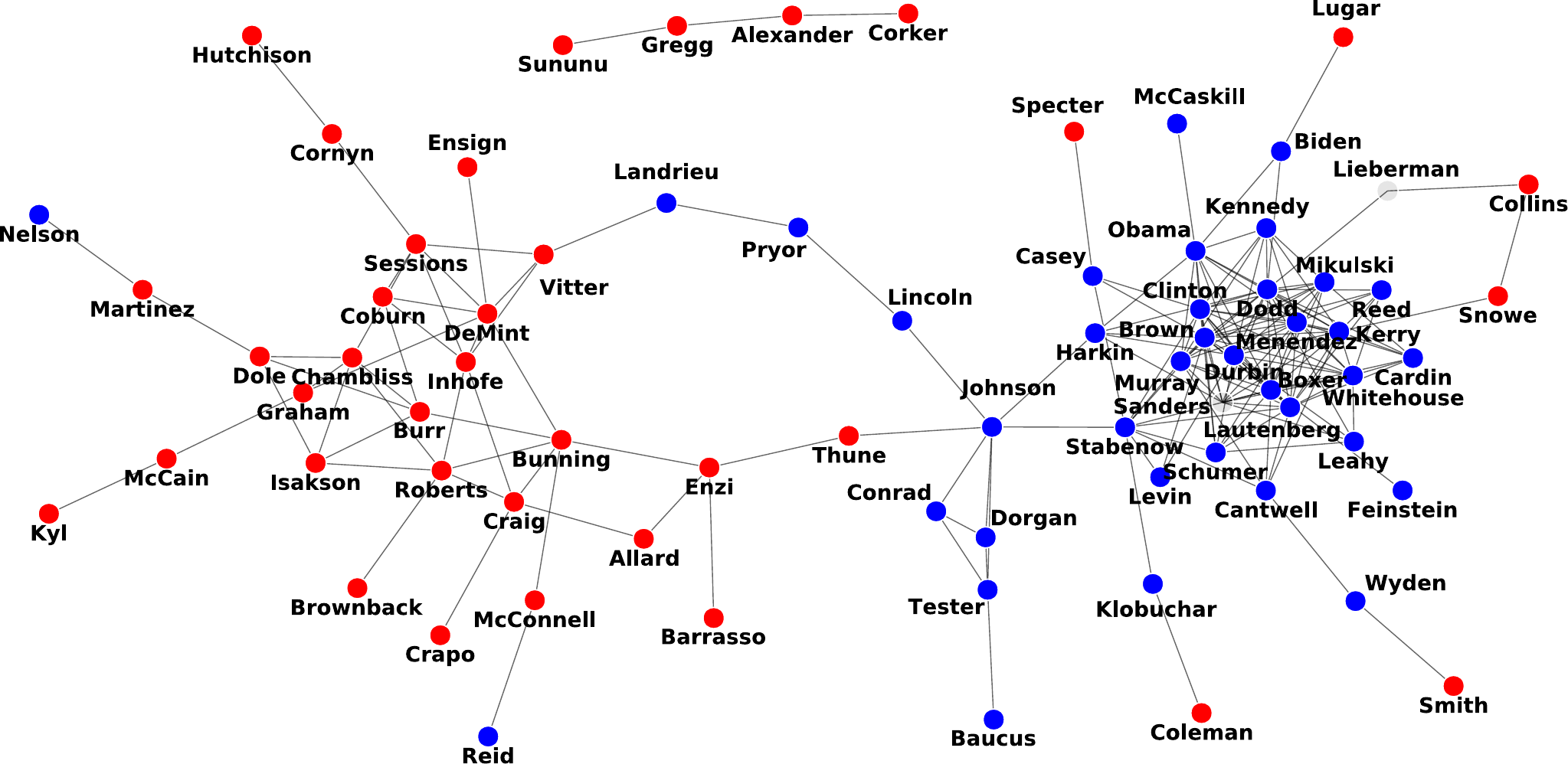}}\caption{{\small{}The two largest connected components of the US senate bill
co-sponsorship network for the 110th congress, pruned down to network
density 2 using a) weight thresholding, b) using our bimodal significance
metric. Here, node colors indicate party membership.\label{fig:senate-cosponsorship-network-both}}}
\end{figure*}

It is not clear whether one can find a closed-form solution for equations
(\ref{eq:logistics1}) and (\ref{eq:logistics2}). However, one can
solve them numerically using iterative methods. %
\begin{comment}
Here I'll use the \textbf{(What's the name of the method?)} method.
\end{comment}
{} In (\ref{eq:logistics1}) and (\ref{eq:logistics2}) we have already
written the system of the equations in the form:
\begin{equation}
x_{i}=\phi_{i}(\left\{ x_{j}\right\} ,\left\{ y_{\beta}\right\} ),\quad y_{\alpha}=\psi_{\alpha}(\left\{ x_{j}\right\} ,\left\{ y_{\beta}\right\} ).
\end{equation}
The solution is therefore the fixed point of the system of transformations
defined by $\phi_{i},i=1,\cdots,m$ and $\psi_{\alpha},\alpha=1,\cdots n.$
In order to compute the fixed point, we start with initial guesses
for each of the $x_{i}$ and $y_{\alpha}$ and iterate the following
equations until convergence:
\begin{equation}
x_{i}^{[k+1]}=\phi_{i}\left(\left\{ x_{j}^{[k]}\right\} ,\left\{ y_{\beta}^{[k]}\right\} \right),
\end{equation}
\begin{equation}
y_{\alpha}^{[k+1]}=\psi_{\alpha}\left(\left\{ x_{j}^{[k]}\right\} ,\left\{ y_{\beta}^{[k]}\right\} \right)
\end{equation}
where the superscript indexes the current step in the iteration. As
the initial values, we use $x_{i}=f_{i}/\sqrt{N}$ and $y_{\alpha}=g_{\alpha}/\sqrt{N}$
which correspond to the first order solution in terms of $x_{i}y_{\alpha}.$
Figure \ref{fig:solution-numerical} shows the results from the numerical
solution of these equations for the US senate cosponsorship data with
$m=3613,$ $n=102,$ such that the co-occurrence graph has 5151 edges.
For details of this data and further discussion, see section \ref{sec:Application-to-data}.

\section{Co-occurrence network}

Having computed the null model's inter-layer connection probabilities,
we now proceed to derive the probability distribution for the co-occurrence
weight of pairs of symbols (first layer nodes). The quantity of interest
is the probability distribution for the random variables $M(s_{i},s_{j})$
defined as follows:
\begin{eqnarray}
M(s_{i},s_{j}): & = & \mbox{number of nodes in the second}\\
 &  & \mbox{layer linked both to \ensuremath{s_{i}} and \ensuremath{s_{j}.}}\nonumber 
\end{eqnarray}
The expectation value of this random variable is given by 
\begin{eqnarray}
\mathbb{E}\left[M(s_{i},s_{j})\right] & = & \sum_{\alpha=1}^{n}p_{i\alpha}p_{j\alpha}
\end{eqnarray}
\begin{comment}
\[
=\sum_{\alpha=1}^{n}\frac{x_{i}x_{j}y_{\alpha}^{2}}{\left(1+x_{i}y_{\alpha}\right)\left(1+x_{j}y_{\alpha}\right)}.
\]
\end{comment}
If $x_{i}\neq x_{j},$ the summand simplifies to 
\begin{eqnarray}
p_{i\alpha}p_{j\alpha} & = & \frac{x_{i}x_{j}y_{\alpha}}{x_{i}-x_{j}}\left[\frac{1}{1+x_{j}y_{\alpha}}-\frac{1}{1+x_{i}y_{\alpha}}\right]
\end{eqnarray}
and thus, the sum over $\alpha$ becomes
\begin{align}
\mathbb{E}\left[M(s_{i},s_{j})\right] & =\sum_{\alpha}p_{i\alpha}p_{j\alpha}=\frac{1}{x_{i}-x_{j}}\left[x_{i}f_{j}-x_{j}f_{i}\right]\nonumber \\
 & \mbox{for }x_{i}\neq x_{j}.
\end{align}
If $x_{i}=x_{j},$ however, this simplification is not valid and we
must compute the full sum
\begin{align}
\mathbb{E}\left[M(s_{i},s_{j})\right] & =\sum_{\alpha}p_{i\alpha}p_{j\alpha}=\sum_{\alpha}\frac{x_{i}^{2}y_{\alpha}^{2}}{\left(1+x_{i}y_{\alpha}\right)^{2}}\nonumber \\
 & \mbox{for}\,x_{i}=x_{j}.
\end{align}

Similarly, we may compute the variance:
\begin{equation}
\operatorname{Var}\left[M(s_{i},s_{j})\right]=\sum_{\alpha}p_{i\alpha}p_{j\alpha}(1-p_{i\alpha}p_{j\alpha}).
\end{equation}
Using these expressions we compute and store $\mathbb{E}\left[M(s_{i},s_{j})\right]$
and $\operatorname{Var}\left[M(s_{i},s_{j})\right]$ once for every
edge in the observed graph. This operation has time complexity $O(n\left|E\right|)$
where $\left|E\right|$ is the size of the edge set of the co-occurence
network.

\section{Distribution of the co-occurrence weights}

The final step is to estimate the probability distribution of $M(s_{i},s_{j})$
so that a \textit{p}-value may be computed for each edge. Note that
$M(s_{i},s_{j})$ is the sum of $n$ binary indicator variables each
indicating whether $s_{i}$ and $s_{j}$ ``co-occurred'' in a set
$u_{\alpha}.$ Therefore, for large $n,$ by the central limit theorem,
we expect the distribution to approach a normal distribution. However,
in general this approximation does not yield accurate results. To
be precise, the sum of independent and different Bernoulli variables
is known as the \textit{Poisson binomial distribution. }Simple closed
form expressions of the pdf and cdf for this distribution aren't known,
but various approximations as well as exact, albeit computationally
expensive numerical estimation methods are known \cite{FernandezPoissonBinom,Hong201341}.
Here we use the so-called \textit{refined normal approximation }(RNA)
due to Volkova \cite{volkova_refinement_1996} which is a modification
of the normal approximation. See Appendix A for details. 

Given the cdf $F_{ij}(k)$ for the null distribution of the weight
between nodes $i,j$ in the co-occurrence graph and an observed weight
$w_{ij}$, we compute the pvalue $\pi_{ij}$ 
\begin{equation}
\pi_{ij}(w_{ij})=1-F_{ij}(w_{ij})
\end{equation}
and define the significance metric as $-\log(\pi_{ij}(w_{ij})).$

\section{Application to data\label{sec:Application-to-data}}

\begin{figure}

\subfloat[]{\includegraphics[width=0.5\columnwidth]{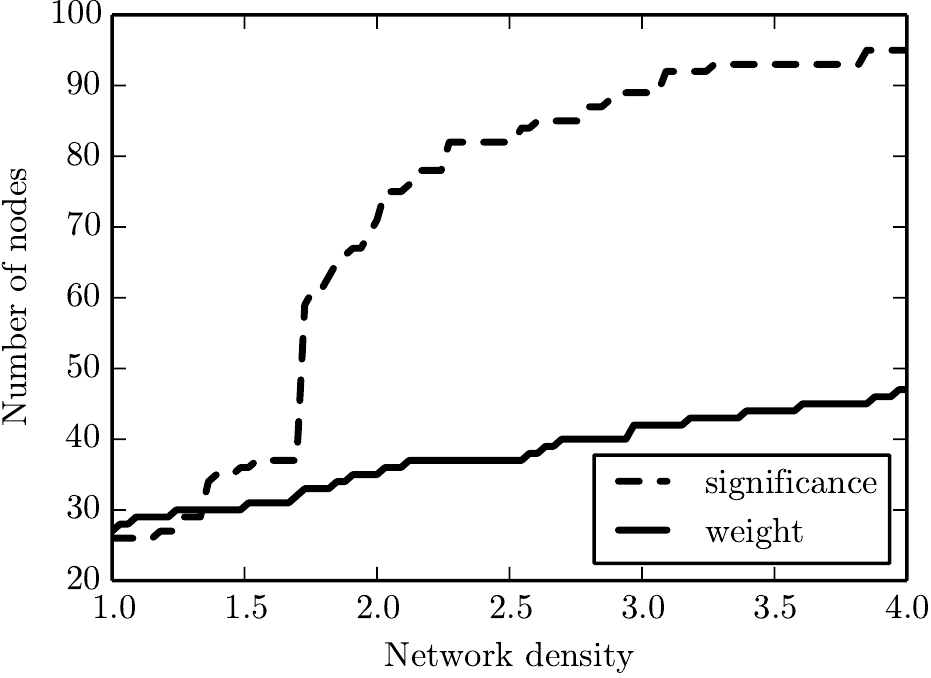}}\subfloat[]{\includegraphics[width=0.5\columnwidth]{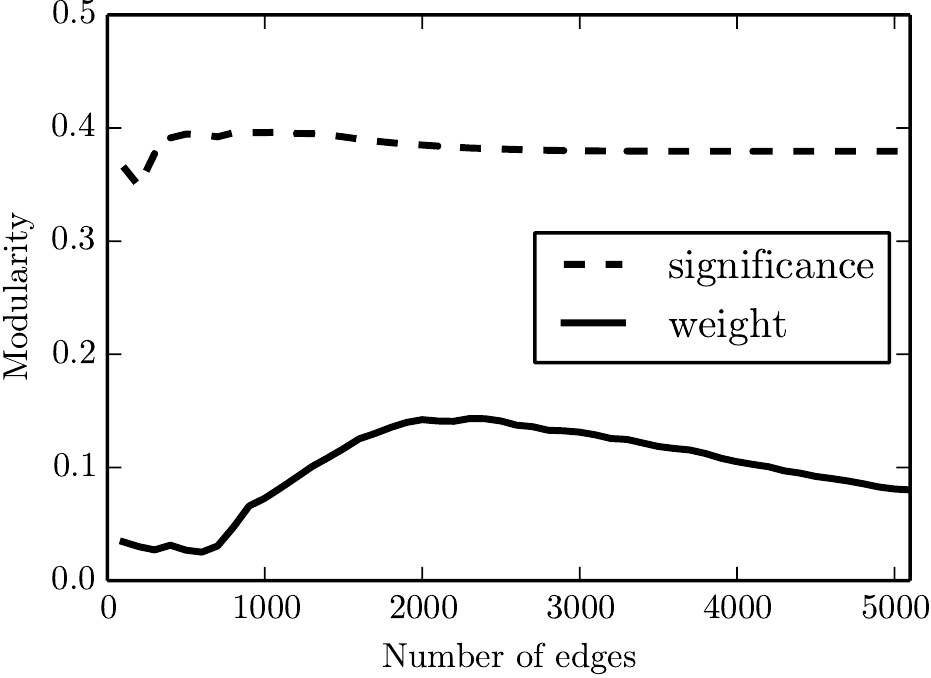}

}\caption{{\small{}a) The size of the giant component for the US senate co-sponsorship
network pruned down to various network densities using the bimodal
significance measure as well as weight thresholding. b) The modularity
of the same networks according to party membership.\label{fig:pruning-comparison-stats}}}

\end{figure}

In this section we present the results of the application of the filter
to the senate bill cosponsorship in the 110th US congress (2007-2008).
The data is from \cite{fowler_connecting_2006,fowler_legislative_2006}
and contains a list of all bills introduced in the senate and for
each one, the list of senators who cosponsored the bill. Aside from
its original sponsor, a bill can also be cosponsored by an arbitrary
number of other senators. Senators cosponsor bills for a variety of
reasons, including partisan allegiance, lending support and forming
strategic alliances, and simply increasing their own visibility and
perceived political clout. Regardless, being cosponsors of a given
bill is a signal of affinity as regards the legislative process. The
data then consists of a bipartite graph where the nodes represent
the senators in the first layer and the bills in the second layer
and an inter-layer link indicates cosponsorship of a bill by a senator.
The co-sponsorhip network is then the projection of this bipartite
graph onto the first layer. The full co-occurence network consists
of 102 nodes and more than 5000 edges, a rather dense graph with no
visible structure. Figure \ref{fig:senate-cosponsorship-network-both}
shows this network pruned using naive weight thresholding as well
as our significance measure. Each graph shows the giant component
as well as the next largest connected component of the graph after
it is pruned down to network density 2 using each pruning scheme.
The graph on the left shows a cluster of mostly Democrats with the
rest of the graph more or less disintegrated. The one on the right
on the other hand, shows most of the nodes connected through the giant
component, which demonstrates a highly modular community structure
reflecting the main partisan division with the senate. Both figures
are rendered using the Kamada-Kawai graph layout, a popular force
directed layout algorithm. Figure \ref{fig:pruning-comparison-stats}
compares weight thresholding and the bimodal filter. On the left,
the size of the giant components truncated at various network densities
are compared. With our significance measure, the giant component already
contains about 80\% of all the nodes at density 2 and nearly all at
density 4, whereas weight thresholding leaves the graph rather disintegrated
up to high densities: a rather small giant component, with the rest
of the nodes scattered across singletons and otherwise very small
components. The figure on the right compares the two methods in terms
of their ability to reveal the partisan divide within the senate.
Given the known party memberships of US senators, we computed the
modularity scores of the pruned graphs at different truncation levels,
both for weight thresholding as well as our bimodal filtering technique.
The modularity of graphs resulting from our filtering technique is
consistently and significantly higher than those produced by weight
thresholding, showing that the partisan divide is manifest much more
clearly with the application of our filter.

\appendix

\section{Refined normal approximation}

In this appendix we describe the refined normal approximation for
the cdf of the Poisson binomial distribution due to Volkova \cite{volkova_refinement_1996}.
For the sum of $n$ independent Bernoulli random variables with means
$p_{i,}\,i=1,2,\cdots n,$ The cdf, $F(k)$ is approximately given
by 
\begin{equation}
F(k)\approx G\left(\frac{k+0.5-\mu}{\sigma}\right),k=0,1,\cdots,n
\end{equation}
where 
\begin{equation}
G(x)=\Phi(x)+\gamma(1-x^{2})\phi(x)/6,
\end{equation}
$\phi(x)$ ,$\Phi(x)$ are the pdf and cdf of the standard normal
distribution respectively and
\begin{equation}
\gamma=\sigma^{-3}\eta\,\,\mbox{where}\,\,\eta=\sum_{j=1}^{n}p_{j}(1-p_{j})(1-2p_{j})
\end{equation}
So, the ingredients necessary for this computation are the following:
\begin{eqnarray}
\mu & = & \sum_{i=1}^{n}p_{i}\\
\sigma^{2} & = & \sum_{i=1}^{n}p_{i}(1-p_{i})\\
\eta & = & \sum_{i=1}^{n}p_{j}(1-p_{j})(1-2p_{j})\\
\phi(x) & = & \frac{1}{\sqrt{2\pi}}e^{-x^{2}/2}\\
\Phi(x) & = & \frac{1}{2}\left[1+\operatorname{erf}\left(\frac{x}{\sqrt{2}}\right)\right]
\end{eqnarray}

\bibliographystyle{ieeetr}
\bibliography{cooccurrence}

\end{document}